\documentclass[sigconf, anonymous=False]{acmart}
\usepackage{tabularx, booktabs}
\usepackage{graphicx}
\usepackage{multirow}
\usepackage{balance}
\usepackage{enumitem}
\usepackage{caption}
\usepackage{diagbox}
\usepackage{booktabs}
\usepackage{subfig}
\usepackage{siunitx}
\usepackage{footmisc}
\usepackage{footnote}
\usepackage[flushleft]{threeparttable}
\usepackage{etoolbox}
\appto\TPTnoteSettings{\footnotesize}

\usepackage{tablefootnote}
\makesavenoteenv{table}
\makesavenoteenv{tabular}

\newcolumntype{C}[1]{>{\centering\arraybackslash}p{#1}}
\newcolumntype{R}[1]{>{\raggedleft\arraybackslash}p{#1}}
\newcolumntype{L}[1]{>{\raggedright\arraybackslash}p{#1}}

\newcommand\mc[1]{\multicolumn{1}{c}{#1}}
\newcommand\mcl[1]{\multicolumn{1}{c!{\color{lightgray}\vrule}}{#1}}

\newcommand\blfootnote[1]{%
\begingroup 
\renewcommand\thefootnote{}\footnote{#1}%
\addtocounter{footnote}{-1}%
\endgroup 
}

\makeatletter
\newcommand{\ssymbol}[1]{^{\@fnsymbol{#1}}}
\makeatother

\makeatletter
\def\hlinew#1{%
  \noalign{\ifnum0=`}\fi\hrule \@height #1 \futurelet
   \reserved@a\@xhline}
\makeatother

\AtBeginDocument{%
  \providecommand\BibTeX{{%
    \normalfont B\kern-0.5em{\scshape i\kern-0.25em b}\kern-0.8em\TeX}}}

\copyrightyear{2021}
\acmYear{2021}
\setcopyright{acmcopyright}\acmConference[CIKM '21]{Proceedings of the 30th ACM International Conference on Information and Knowledge Management}{November 1--5, 2021}{Virtual Event, QLD, Australia}
\acmBooktitle{Proceedings of the 30th ACM International Conference on Information and Knowledge Management (CIKM '21), November 1--5, 2021, Virtual Event, QLD, Australia}
\acmPrice{15.00}
\acmDOI{10.1145/3459637.3482358}
\acmISBN{978-1-4503-8446-9/21/11}
\begin{document}
\fancyhead{}

\title{Jointly Optimizing Query Encoder and Product Quantization to Improve Retrieval Performance}


%
%
%
%
%

\author{Jingtao Zhan$^{1}$, Jiaxin Mao$^{2}$, Yiqun Liu$^{1\star}$, Jiafeng Guo$^{3}$, Min Zhang$^{1}$, Shaoping Ma$^{1}$}
\affiliation{%
  \institution{${1}$ Department of Computer Science and Technology, Institute for Artificial Intelligence, \\
  Beijing National Research Center for Information Science and Technology, Tsinghua University, Beijing 100084, China
  }
  }
  
\affiliation{%
  \institution{${2}$ Beijing Key Laboratory of Big Data Management and Analysis Methods, Gaoling School of Artificial Intelligence, \\ Renmin University of China, Beijing 100872, China
}
  }

\affiliation{%
  \institution{${3}$ CAS Key Lab of Network Data Science and Technology, Institute of Computing Technology, \\
  		Chinese Academy of Sciences, Beijing, China
  }
  }

\email{jingtaozhan@gmail.com, maojiaxin@gmail.com, yiqunliu@tsinghua.edu.cn, guojiafeng@ict.ac.cn}

\renewcommand{\shortauthors}{Zhan, et al.}

\begin{abstract}
Recently, Information Retrieval community has witnessed fast-paced advances in Dense Retrieval~(DR), which performs first-stage retrieval with embedding-based search. Despite the impressive ranking performance, previous studies usually adopt brute-force search to acquire candidates, which is prohibitive in practical Web search scenarios due to its tremendous memory usage and time cost. To overcome these problems, vector compression methods have been adopted in many practical embedding-based retrieval applications. One of the most popular methods is Product Quantization~(PQ). However, although existing vector compression methods including PQ can help improve the efficiency of DR, they incur severely decayed retrieval performance due to the separation between encoding and compression. To tackle this problem, we present JPQ, which stands for Joint optimization of query encoding and Product Quantization. It trains the query encoder and PQ index jointly in an end-to-end manner based on three optimization strategies, namely ranking-oriented loss, PQ centroid optimization, and end-to-end negative sampling. We evaluate JPQ on two publicly available retrieval benchmarks. Experimental results show that JPQ significantly outperforms popular vector compression methods. Compared with previous DR models that use brute-force search, JPQ almost matches the best retrieval performance with 30x compression on index size. The compressed index further brings 10x speedup on CPU and 2x speedup on GPU in query latency.
\end{abstract}

\ccsdesc[500]{Information systems~Retrieval models and ranking}
\ccsdesc[500]{Information systems~Search index compression}
\ccsdesc[300]{Information systems~Information retrieval}

\keywords{dense retrieval; index compression; neural ranking}

\maketitle

\blfootnote{$^\star$Corresponding author}

\section{Introduction}

Traditional web search relies on Bag-of-Words~(BoW) retrieval models like BM25~\cite{robertson1994some} for efficient first-stage retrieval. Recently, with the rapid progress in representation learning~\cite{bengio2013representation} and deep pre-trained language models~\cite{vaswani2017attention, devlin2019bert, liu2019roberta}, Dense Retrieval~(DR)~\cite{cai2021semantic} has become a popular paradigm to improve retrieval performance.
In this paradigm, dual-encoders are employed to embed user queries and documents in a latent vector space.
To generate result candidates based on embedded queries and documents in the space, most existing DR models~\cite{ding2020rocketqa, hofstatter2021efficiently, xiong2020approximate, zhan2020repbert} rely on brute-force search. They significantly outperform BoW models in terms of effectiveness and benefit downstream tasks like OpenQA~\cite{guu2020realm, karpukhin2020dense}. However, brute-force search is not suitable for practical Web search scenarios due to its high computational cost. For online services, most solutions rely on Approximate Nearest Neighbor Search~(ANNS) to perform efficient vector search~\cite{zhang2021joint, cao2016deep}.

As an important implementation of ANNS, vector compression methods~\cite{jegou2010product, ge2013optimized} have been adopted in many practical embedding-based retrieval applications for the following two reasons. 
Firstly, the size of the uncompressed embedding index is very large. For example, it is typically an order of magnitude larger than the traditional BoW index in existing DR studies~\cite{zhan2020optimizing, xiong2020approximate, zhan2020repbert}. 
Secondly, the embedding index is usually required to be loaded into system memory or even GPU memory~\cite{johnson2019billion}, whose size is highly limited.
Therefore, compressing the embedding index is necessary when DR is applied in practical Web search scenario.
Popular compression methods include Product Quantization~(PQ)~\cite{jegou2010product, ge2013optimized} and Locality Sensitive Hashing~(LSH)~\cite{indyk1998approximate}.

Although existing vector compression methods can help improve efficiency for DR models, they suffer from a few drawbacks in practical scenarios. 
Many popular compression methods~\cite{jegou2010product, ge2013optimized, guo2020accelerating} cannot benefit from supervised information because they use the task-independent reconstruction error as the loss function in training. Besides, the encoders and the compressed index are separately trained and thus may not be optimally compatible.
Recently, some studies~\cite{chen2020differentiable, zhang2021joint, cao2016deep, yu2018product} jointly train the encoders and the compressed index. 
However, these studies are not designed for Web search scenarios and do not take into account the characteristics of DR processes, such as the importance of negative sampling~\cite{zhan2020optimizing, xiong2020approximate}.
Moreover, they still attempt to minimize the reconstruction error rather than to optimize retrieval performance directly~\cite{chen2020differentiable, zhang2021joint, cao2016deep}.
In our experiments described in Section~\ref{sec:exp_compare_anns}, we find the existing methods still lead to a significant compromise in ranking performance.  

\begin{figure}
    \includegraphics[width=0.9\linewidth, keepaspectratio=True]{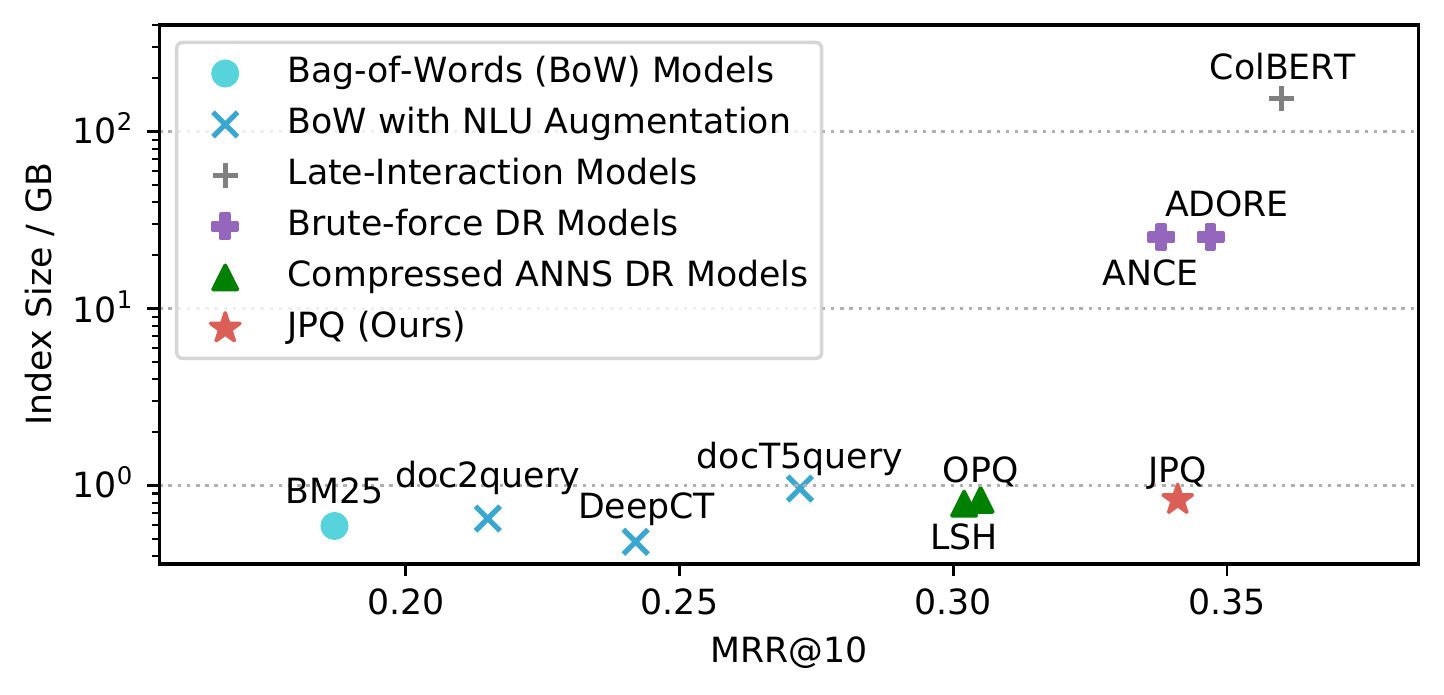}
    \caption{Effectiveness~(MRR@10) versus Index Size~(log-scale) for different retrieval methods on MS MARCO Passage Ranking~\cite{bajaj2016ms}. The index size of JPQ is only 1/186 of the size of ColBERT.
    } 
    \label{fig:ms_passage_all_models}
    \vspace{-3mm}
\end{figure}

In Figure~\ref{fig:ms_passage_all_models}, we show the trade-offs between index size and ranking performance for a variety of existing first-stage retrieval models including Brute-force DR Models~(before compression), Compressed ANNS DR Models~(after compression), BoW models, and Late Interaction models.
We report retrieval effectiveness (MRR@10) as well as the index size (log-scale) on MS MARCO Passage Ranking~\cite{bajaj2016ms}, a widely-adopted benchmark dataset.
As the figure shows, although dual-encoders improve the search performance compared with BoW models, they lead to a significant increase in the index size. 
However, when the index is compressed by popular methods like LSH~\cite{indyk1998approximate} and OPQ~\cite{ge2013optimized}, the retrieval effectiveness is hurt by a large margin. 

To maintain retrieval effectiveness while reducing index sizes using compression, we propose JPQ, which stands for Joint optimization of query encoding and Product Quantization\footnote{Code and trained models are available at \url{https://github.com/jingtaozhan/JPQ}.}.
It aims to optimize the ranking performance in an end-to-end manner instead of following the existing encoding-compression two-step procedure.
To achieve this goal, it jointly optimizes the query encoder and PQ index with three strategies:
1) We use \emph{ranking-oriented loss} for JPQ, abandoning the partial ranking loss output by dual-encoders alone~\cite{zhan2020optimizing, xiong2020approximate, zhan2020repbert} and the task-independent reconstruction loss widely-used for training PQ~\cite{chen2020differentiable, zhang2021joint, cao2016deep}. It is computed in an end-to-end manner, i.e., the actual loss produced by dual-encoders with PQ index, and thus can evaluate the ranking performance accurately.
2) Training PQ index with \emph{ranking-oriented loss} is non-trivial due to problems like differentiability and overfitting. To tackle these problems, JPQ proposes to use \emph{PQ centroid optimization}, which only trains a small but crucial number of PQ parameters, i.e., PQ Centroid Embeddings. Other PQ parameters are well initialized and fixed during training.
3) Besides the \emph{ranking-oriented loss}, JPQ uses \emph{end-to-end negative sampling} to further improve end-to-end ranking performance. Given several training queries, it performs end-to-end retrieval using current encoder and PQ parameters in training. The top-ranked irrelevant documents are utilized as negatives. Through penalizing the scores of these documents, JPQ learns to improve end-to-end ranking performance. 

To verify the effectiveness and efficiency of JPQ, we conduct extensive experiments on two publicly available benchmarks~\cite{bajaj2016ms, craswell2020overview} and compare JPQ against a wide range of existing ANNS methods and first-stage retrieval models. Experimental results show that: 
1) JPQ substantially reduces the index size and \emph{improves the retrieval efficiency without significantly hurting the retrieval effectiveness}. Therefore, compared with existing index compression methods that trade effectiveness for efficiency, JPQ outperforms them in terms of ranking effectiveness by a large margin; compared with non-exhaustive ANNS methods that do not compress the index, JPQ achieves a better ranking performance with a 30x smaller index file. 
2) JPQ outperforms existing first-stage retrieval approaches in terms of effectiveness and efficiency. JPQ is much more effective than the BoW models with a similar index size. It is more efficient than existing DR models that use brute-force search. Specifically, It gains similar ranking performance with state-of-the-art DR models while providing 30x index compression ratio, 10x CPU speedup, and 2x GPU speedup. Compared with late-interaction models, JPQ gains similar recall with a several orders of magnitude smaller index and 5x GPU speedup. 
We also conduct an ablation study for JPQ on the passage ranking dataset~\cite{bajaj2016ms}. According to the experimental results, all three strategies, namely ranking-oriented loss, PQ centroid optimization, and end-to-end negative sampling, contribute to its effectiveness.

\section{Background and Related Works}

This section introduces the background and related works.
Several commonly used notations are:  $\mathcal{C}$ denotes the set of all documents; $N$ is the number of all documents; $n$ is the number of documents returned by the retrieval algorithms; and $D$ is the embedding dimension.


\subsection{Dual-Encoders}
Dual-encoders, i.e., the query encoder and the document encoder, represent queries and documents with embeddings.
Let $f$ be the dual-encoders, which takes the input of a query $q$ or a document $d$ and outputs an embedding, $\vec{q}$ or $\vec{d}$:
\begin{equation}
	\vec{q} = f(q) \in \mathbb{R}^D \quad \vec{d} = f(d) \in \mathbb{R}^D
\end{equation}
Let $ \langle , \rangle$ be the embedding similarity function. The predicted relevance score $s(q,d)$ is: 
\begin{equation}
s(q, d) = \langle \vec{q}, \vec{d} \rangle 
\end{equation}

Dual-encoders are mainly trained by negative sampling methods~\cite{zhan2020repbert, xiong2020approximate, ding2020rocketqa}. 
Some studies use all irrelevant documents~\cite{huang2020embedding, ding2020rocketqa} as negatives. Some studies use hard negatives retrieved by BM25~\cite{zhan2020repbert, gao2020complementing} or a warm-up DR model~\cite{xiong2020approximate}. 
Recently, \citet{zhan2020optimizing} propose dynamic hard negative sampling and show that hard negatives help improve top ranking performance.

\subsection{Brute-force Search}
Brute-force search is utilized by many previous DR studies~\cite{zhan2020optimizing, xiong2020approximate, hofstatter2021efficiently} to retrieve candidates. We briefly formulate the search process and analyze its efficiency. 

\subsubsection{Search Procedure} 
The search procedure contains two stages, score computation and sorting. Given a query embedding $\vec{q}$, it computes the relevance score $s(q,d)$ using $\langle \vec{q}, \vec{d} \rangle $ for each document $d \in \mathcal{C}$. Then, it sorts all documents based on $s(q,d)$ and returns the top-$n$ documents. We formally express the procedure as follows.
\begin{equation}
{\rm results} = {\rm sort}(d \in \mathcal{C} \ {\rm based \ on}\ s(q,d))[:n]  
\end{equation}

\subsubsection{Time Complexity} \mbox{}
For score computation, the complexity is $O(ND)$. 
For sorting, the complexity is $O(N {\rm log}n)$.
Therefore, the overall time complexity is $O(ND + N {\rm log}n)$.

\subsubsection{Index Size} \mbox{}
The index is actually all document embeddings. 
Since each float consumes $4$ bytes, the index size is $4ND$ bytes.

\subsection{ANNS}
\label{sec:background_anns}

ANNS achieves highly efficient vector search by allowing a small number of errors.
Generally, there are two kinds of ANNS algorithms. One is non-exhaustive ANNS methods, and the other is vector compression methods.
Both are important and usually combined in practice. 

Non-exhaustive ANNS methods do not compress the index. They reduce the number of candidates for each query to speed up the retrieval process. 
Formally, let $\Omega$ be one method, and $\Omega(q, \mathcal{C})$ be the returned candidates for query $q$. The search process is:
\begin{equation}
{\rm results} = {\rm sort}(d \in \Omega(q,\mathcal{C}) \ {\rm based \ on}\ s(q,d))[:n] \\  
\end{equation}
Popular algorithms include tree-based methods~\cite{muja2014scalable, Github:annoy}, inverted file methods~\cite{babenko2014inverted, jegou2010product}, and graph-based methods~\cite{malkov2018efficient}.


Vector compression methods mainly aim to compress the index but are also able to accelerate retrieval.
They learn a new score function $s^\dagger$, which is fast to compute and requires a small document index. We formulate the search process as follows:
\begin{equation}
{\rm results} = {\rm sort}(d \in \mathcal{C} \ {\rm based \ on}\ s^\dagger(q,d))[:n] \\  
\end{equation}
Popular algorithms include hashing~\cite{indyk1998approximate} and quantization~\cite{jegou2010product, ge2013optimized}.



\subsection{Product Quantization}
\label{sec:background_pq}

Product Quantization~(PQ)~\cite{jegou2010product} is a popular ANNS algorithm and belongs to the second kind of ANNS introduced in the previous section. 

\subsubsection{Approximate Score Function} \mbox{}

PQ replaces $s$ with a new approximate score function $s^\dagger$. It quantizes the document embedding $\vec{d} \in \mathbb{R}^D$ to $\vec{d}^\dagger \in \mathbb{R}^D$. Then it uses $\vec{d}^\dagger$ to compute similarity:
\begin{equation}
\label{eq:pq_score_brute}
	s^\dagger(q,d) = \langle \vec{q}, \vec{d}^\dagger \rangle 
\end{equation}
Next, we introduce how it quantizes document embeddings.

\subsubsection{Quantizing Document Embeddings} \mbox{}

PQ defines $M$ sets of embeddings, each of which includes $K$ embeddings of dimension $D/M$. We call them PQ Centroid Embeddings. To quantize a document embedding, PQ picks one from each of these $M$ sets and then concatenates the $M$ picked embeddings as the quantized document embeddings. 

Formally, let $\vec{c}_{i,j}$ be the $j_{th}$ centroid embedding from the $i_{th}$ set:
\begin{equation}
	\vec{c}_{i,j} \in \mathbb{R}^\frac{D}{M} \quad (1 \leq i \leq M, 1 \leq j \leq K) 
\end{equation}
Given a document embedding $\vec{d}$, PQ picks the $\varphi_i(d)$$_{th}$ centroid embedding from the $i_{th}$ centroid set ($1 \leq i \leq M$). Then it concatenates the $M$ picked embeddings as $\vec{d}^\dagger$:
\begin{equation}
	\vec{d} \rightarrow \vec{d}^\dagger = \vec{c}_{1,\varphi_1(d)},\vec{c}_{2,\varphi_2(d)},...,\vec{c}_{M,\varphi_M(d)} \in \mathbb{R}^D
\end{equation}
where comma denotes the concatenation operation.
We call $\varphi_i(d)$ Index Assignments.


\subsubsection{Optimization Objective} \mbox{}

PQ trains $\{\vec{c}_{i,j}\}$ and $\varphi$ to minimize the MSE loss between the original document embedding $\vec{d}$ and the quantized one $\vec{d}^\dagger$. 
\begin{equation}
\label{eq:pq_optimize_objective}
	\{\vec{c}_{i,j}\}, \varphi = \arg \min \lVert \vec{d} - \vec{d}^\dagger \rVert ^2
\end{equation}
The loss is also called reconstruction error. In this way, $s^\dagger(q,d)$ approximates the true $s(q,d)$.


\subsubsection{Search Procedure}\mbox{}
\label{sec:background_pq_search_process}

PQ uses an equivalent but very efficient way to compute Eq.~(\ref{eq:pq_score_brute}). It firstly splits the query embedding equally to $M$ sub-vectors:
\begin{equation}
	\label{eq:pq_split_query_emb}
	 \vec{q} = \vec{q}_1,\vec{q}_2,...,\vec{q}_M
\end{equation}
where $\vec{q}_i$ denotes the $i_{th}$ sub-vector and is of dimension $D/M$. Then it computes the similarity score between the query sub-vectors and PQ Centroid Embeddings: 
\begin{equation}
	\tau_{i,j} = \langle \vec{q}_i, \vec{c}_{i,j} \rangle \quad (1 \leq i \leq M, 1 \leq j \leq K) 
\end{equation}
After constructing a lookup table with $\tau_{i,j}$, PQ efficiently computes $s^\dagger(q,d)$. For example, if $\langle, \rangle$ is inner product, PQ only needs to sum corresponding $\tau_{i,j}$:
\begin{equation}
\label{eq:pq_aggregate_score}
	s^\dagger(q,d) =\sum_{i=1}^M \tau_{i, \varphi_i(d)}  
\end{equation}

\subsubsection{Time Complexity} \mbox{}

The major time cost is computing Eq.~(\ref{eq:pq_aggregate_score}) for all documents and sorting.
The overall time complexity is $O(NM+N{\rm log}n)$. Compared with brute-force search, the speedup ratio is $(D+{\rm log}n)/(M+{\rm log}n)$. 

\subsubsection{Index Size} \mbox{}

PQ does not explicitly store $\vec{d}^\dagger$. Instead, it only stores the PQ Centroid Embeddings $\{\vec{c}_{i,j}\}$ and Index Assignments $\{\varphi_i(d)\}$.
$K$ usually equals to or is less than 256 so that $\varphi_i(d)$ can be stored with one byte.
The overall index size is $4KD+NM \approx NM$ bytes. 
Compared with brute-force search, the compression ratio is $4D/M$.

\subsection{PQ Variants}

There are several variants of PQ. One of the most popular variants is OPQ~\cite{ge2013optimized}, which adds a linear transformation before quantization.  
Some variants~\cite{babenko2014additive, martinez2018lsq++} use different architectures to better minimize the reconstruction error, while training and searching is more complex.
This paper leaves joint optimizing with these methods for future work.
Recently, ~\citet{guo2020accelerating} propose ScaNN, which uses a new loss to optimize PQ parameters and achieves state-of-the-art results on ANN benchmarks.

Inspired by the progress of deep learning, some deep methods~\cite{chen2020differentiable, cao2016deep, klein2019end, zhang2021joint} are proposed to perform the feature learning and compression simultaneously.
However, they focus on other areas and do not consider the characteristics of DR in document retrieval, such as the importance of negative sampling~\cite{zhan2020optimizing, xiong2020approximate}.
The compressed representations are trained based on the reconstruction error rather than directly based on the supervised signals~\cite{chen2020differentiable, zhang2021joint, cao2016deep}.
We implement a baseline based on ~\citet{chen2020differentiable} and \citet{zhang2021joint}. Section~\ref{sec:exp_compare_anns} will show it still causes significant performance loss after compression.

\section{The JPQ Model}
We propose JPQ, Joint optimization of query encoding and Product Quantization. 
In the following, we will outline the overall architecture, describe the three strategies, summarize the optimization objective, and analyze its efficiency.

\subsection{Overall Architecture}


\begin{figure}
    \subfloat[Workflow of existing training methods.]{\label{fig:baseline_train}\includegraphics[width=0.85\linewidth]{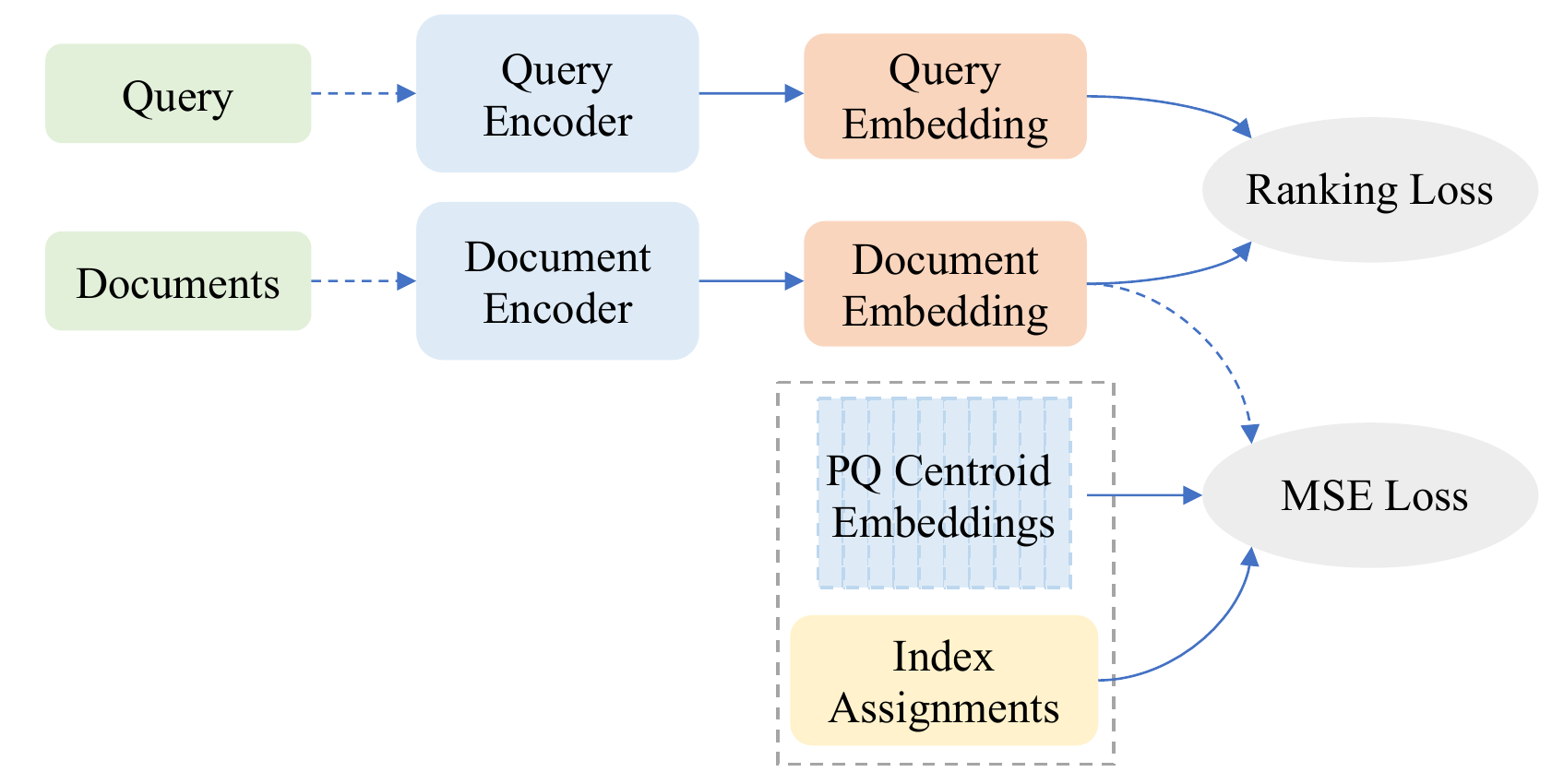}} \\
    \subfloat[Workflow of JPQ.]{\label{fig:our_train}\includegraphics[width=0.85\linewidth]{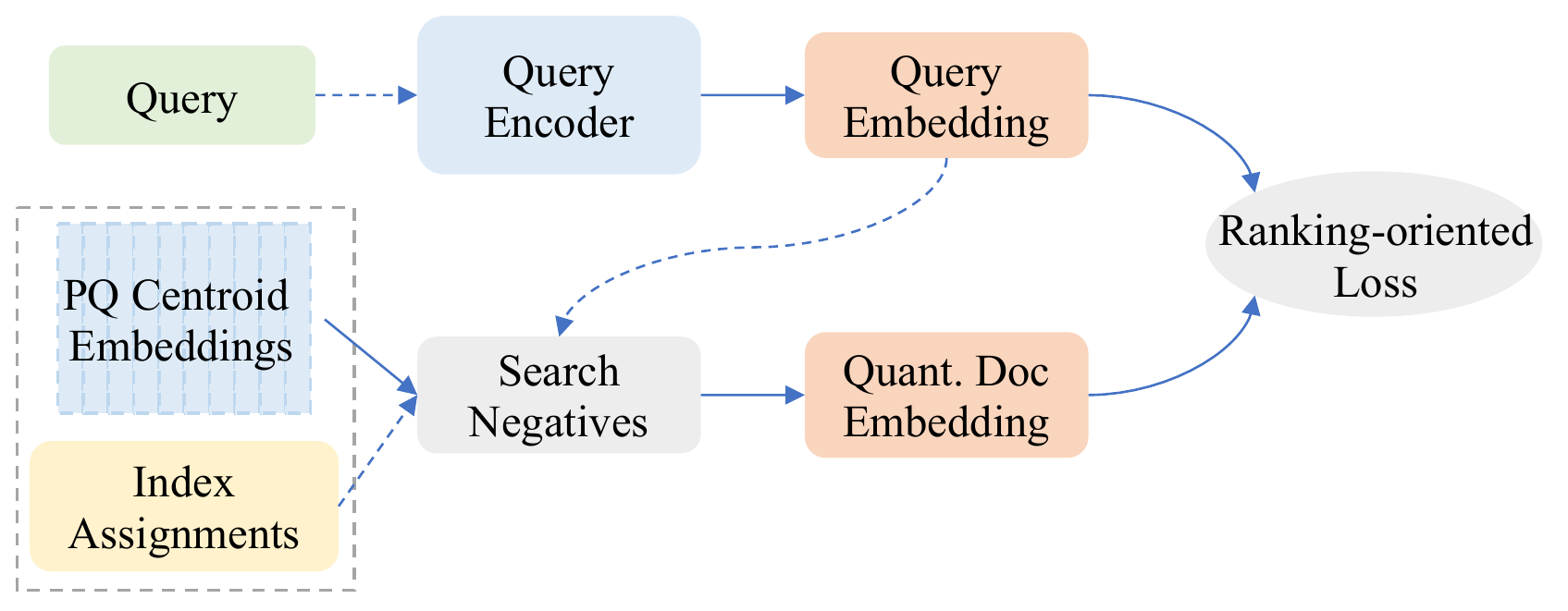}}
    \caption{Training workflows. Solid arrows indicate the gradients are backpropagated, whereas the dotted arrows indicate otherwise.}
    \label{fig:overall_previous_and_jpq}
    \vspace{-3mm}
\end{figure}

Figure~\ref{fig:overall_previous_and_jpq} contrasts JPQ with existing training methods. Solid arrows indicate the gradients are backpropagated, whereas the dotted arrows indicate otherwise.

On the top, Figure~\ref{fig:baseline_train} illustrates previous methods~\cite{ge2013optimized, jegou2010product, guo2020accelerating}, which follow an encoding-compression two-step procedure. They firstly train the dual-encoders with ranking loss, which does not consider PQ compression. Then they use the trained document encoder to encode all documents. Given all document embeddings, they finally train the PQ index to minimize the MSE loss~(aka reconstruction error), which is task-independent and cannot benefit PQ index with supervised information.

On the bottom, Figure~\ref{fig:our_train} visualizes the training process of JPQ, which follows an end-to-end paradigm with three strategies:
\begin{itemize}
	\item JPQ proposes to use a new \emph{ranking-oriented loss}, which is the actual loss produced by dual-encoders and PQ index. To compute it, JPQ firstly reconstructs the quantized document embeddings, then computes the actual relevance scores used by PQ for ranking, and finally passes the scores to a pair-wise loss function.
	\item Training PQ index with \emph{ranking-oriented loss} is non-trivial due to problems like differentiability and overfitting. Hence, JPQ proposes \emph{PQ centroid optimization} to address these problems. It initializes Index Assignments and only updates PQ Centroid Embeddings using gradient decent.
	\item JPQ utilizes \emph{end-to-end negative sampling} to further improve ranking performance. The training query embeddings are fed into the `Search Negatives' module, which uses the current PQ parameters to retrieve top-ranked irrelevant documents as negatives.
\end{itemize}  

Next, we will describe the three strategies in detail.

\subsection{Ranking-oriented Loss}
JPQ uses ranking-oriented loss to accurately evaluate the end-to-end ranking performance.
It is computed based on the actual scores utilized for ranking. We elaborate on it in the following.

Previous DR studies~\cite{luan2020sparsedense, zhan2020repbert, gao2020complementing} usually adopt the following pair-wise ranking loss:
\begin{equation}
	\ell(s(q,d^+),s(q,d^-)) 
	\label{eq:previous_loss_func}
\end{equation}
where $s(q,d)$ is the original relevance score output by the dual-encoders, and $\ell: \mathbb{R}\times \mathbb{R} \rightarrow \mathbb{R}$ is a pair-wise loss function.
Nevertheless, it is not the actual scores utilized by many ANNS algorithms during ranking. 
As for PQ, it quantizes documents and uses a new score $s^\dagger(q,d)$ for ranking. 
Since ranking with $s(q,d)$ or $s^\dagger(q,d)$ is likely to yield different ranking lists, training with the above loss cannot effectively improve the actual PQ ranking performance. 

To solve this problem, we use $s^\dagger(q,d)$ to compute the ranking-oriented loss. Since $s^\dagger(q,d)$ equals to the similarity between query embedding and quantized document embeddings, JPQ firstly reconstructs the quantized document embeddings $\vec{d}^\dagger$ from the PQ index:
\begin{equation}
	\vec{d}^\dagger = \vec{c}_{1,\varphi_1(d)},\vec{c}_{2,\varphi_2(d)},...,\vec{c}_{M,\varphi_M(d)}
\end{equation}
Secondly, JPQ computes $s^\dagger(q,d)$ using $\vec{q}$ and $\vec{d}^\dagger$:
\begin{equation}
	s^\dagger(q,d) = \langle \vec{q}, \vec{d}^\dagger \rangle 
\end{equation}
Finally, JPQ passes $s^\dagger(q,d)$ to the pair-wise loss function: 
\begin{equation}
\label{eq:jpq_loss_func}
\ell(s^\dagger(q,d^+),s^\dagger(q,d^-))  
\end{equation}
Because Eq.~(\ref{eq:jpq_loss_func}) utilizes the actual ranking scores, it evaluates the ranking performance accurately and helps improve it using gradient descent.

\subsection{PQ Centroid Optimization}
Training PQ with ranking-oriented loss is non-trivial due to the following two problems related to Index Assignments. Firstly, Index Assignments are not differentiable with respect to the ranking-oriented loss. As described in Section~\ref{sec:background_pq_search_process}, the Index Assignments are utilized to select corresponding PQ Centroid Embeddings, which is hard to optimize using gradient decent. Secondly, even if some methods other than gradient decent can be used to update Index Assignments, e.g., exhaustive enumeration, it may cause overfitting problems because the number of Index Assignments is huge, i.e., proportional to the size of the corpus. 

To resolve these problems, JPQ proposes PQ centroid optimization, which initializes Index Assignments using Figure~\ref{fig:baseline_train} and only trains a very small but crucial number of PQ parameters, PQ Centroid Embeddings. 
These centroid embeddings are differentiable, and it is unlikely to overfit the training data due to its small number. 

Now we illustrate how PQ Centroid Embeddings are updated using gradient decent. We compute its gradients with respect to the ranking-oriented loss in three steps.
Firstly, for many commonly used pair-wise loss, such as hinge loss, RankNet~\cite{burges2005learning}, and LambdaRank~\cite{burges2010ranknet}, we can define $\alpha \geq 0$ as follows:
\begin{equation}
\alpha = -\frac{\partial \ell(s^\dagger(q,d^+),s^\dagger(q,d^-))}{\partial s^\dagger(q,d^+)}  = \frac{\partial \ell(s^\dagger(q,d^+),s^\dagger(q,d^-))}{\partial s^\dagger(q,d^-)} \geq 0  \\
\end{equation}
Secondly, the gradient of $\vec{c}_{i,j}$ with respect to the score $s^\dagger(q,d)$ is:
\begin{equation}
  \frac{\partial s^\dagger(q,d)}{\partial \vec{c}_{i,j}} = \begin{cases}
    \vec{q}_i, & \text{if $j=\varphi_i(d)$}.\\
    0, & \text{otherwise}.
  \end{cases}
\end{equation}
where $\vec{q}_i$ is the $i_{th}$ query embedding sub-vector described in Eq.~(\ref{eq:pq_split_query_emb}). 
Finally, with the above two equations, we can use \emph{chain rule} to derive the gradients of PQ Centroid Embeddings:
\begin{equation}
  \frac{\partial \ell(s^\dagger(q,d^+),s^\dagger(q,d^-))}{\partial \vec{c}_{i,j}} =\begin{cases}
    -\alpha  \vec{q}_i, & \text{if $j=\varphi_i(d^+)$, $j \neq \varphi_i(d^-)$}.\\
    \alpha  \vec{q}_i, & \text{if $j\neq \varphi_i(d^+)$, $j = \varphi_i(d^-)$}.\\
    0, & \text{if $j=\varphi_i(d^+) $, $j= \varphi_i(d^-)$}.\\
    0, & \text{if $j\neq\varphi_i(d^+)$, $j \neq \varphi_i(d^-)$}.
  \end{cases}
\end{equation}

According to the above gradients, PQ centroid optimization has two advantages, benefiting PQ index with supervised signals and helping it evolve with the query encoder.
Firstly, the PQ index benefits from supervised signals through relevance labels and $\alpha$. Relevance labels control the updating conditions. $\alpha$ decides the updating weight, which is small if the ranking is correct and large otherwise.
Secondly, PQ parameters directly evolve with the query encoder through $\vec{q}_i$, which is part of the query encoder's output.

\subsection{End-to-End Negative Sampling}
Besides ranking-oriented loss, JPQ proposes end-to-end negative sampling to further improve end-to-end ranking performance. Note that negative sampling has been shown to be important for training dual-encoders~\cite{zhan2020optimizing, xiong2020approximate}. 

Our intuition is to penalize those top-ranked irrelevant documents and disregard others. The top-ranked negatives greatly affect the ranking performance, whereas low-ranked documents are mostly ignored by the truncated evaluation metric. By penalizing the top-ranked negatives, we can improve the top-ranking performance, which is the target of many popular IR systems, such as Web search engines.

The top-ranked negatives are acquired by real-time end-to-end retrieval at each training step.
As shown in Figure~\ref{fig:our_train}, the training query embeddings are passed to the `Search Negatives' module, which uses the current PQ parameters to retrieve top-$\hat{n}$ irrelevant documents as negatives. Let ${\mathcal{D}^-_q}^\dagger$ be the retrieved negatives and ${\mathcal{D}^+_q}$ be the labeled relevant documents. ${\mathcal{D}^-_q}^\dagger$ is formulated as:
\begin{equation}
{\mathcal{D}^-_q}^\dagger = {\rm sort}(d \in \mathcal{C}\backslash \mathcal{D}^+_q \ {\rm based \ on}\ s^\dagger(q,d))[:\hat{n}] \\  
\end{equation}
Using ${\mathcal{D}^-_q}^\dagger$ as negatives can also be regarded as minimizing the top-$\hat{n}$ pairwise errors, which is in line with the truncated evaluation metric.

\begin{figure}
    \includegraphics[width=0.45\linewidth, keepaspectratio=True]{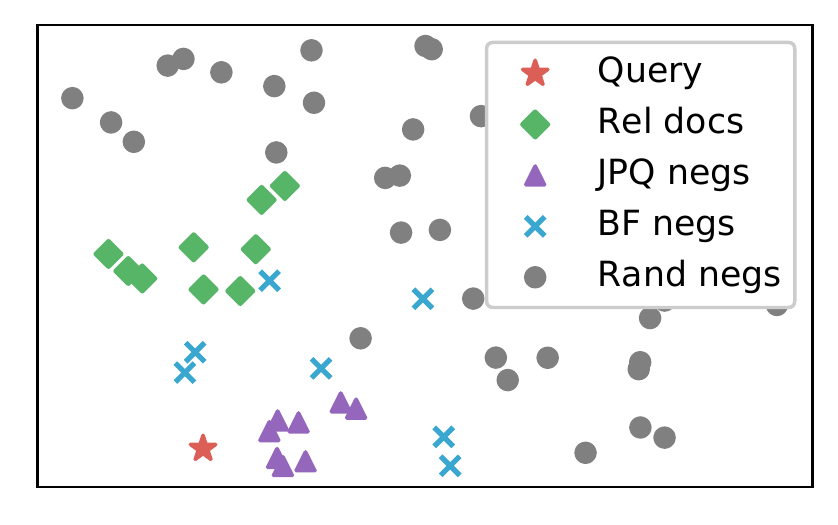}
    \caption{The t-SNE plot of `BF negs'~\cite{zhan2020optimizing} and `JPQ negs' negative sampling methods. The QID is 443396 from TREC DL Track~\cite{craswell2020overview}. 
    } 
    \label{fig:neg_sample}
    \vspace{-3mm}
\end{figure}

Although using top-ranked irrelevant documents as negatives has recently been explored by \citet{zhan2020optimizing}, they utilize brute-force search to retrieve negatives. 
We use `BF negs' and `JPQ negs' to denote their selected negatives and ours, respectively. 
Figure~\ref{fig:neg_sample} illustrates an example from TREC DL 2019 dataset~\cite{craswell2020overview} using t-SNE~\cite{maaten2008visualizing}. According to our intuition, the nearest neighbors to the query should be utilized as negatives. 
As the figure shows, our method selects the actual nearest neighbors owing to the end-to-end retrieval, whereas `BF negs' do not due to the difference between PQ ranking and brute-force search.

\subsection{Optimization Objective}
Now we summarize the optimization objective of JPQ. JPQ uses end-to-end negative sampling and computes the ranking-oriented loss to jointly optimize the query encoder and PQ Centroid Embeddings. Therefore, the optimization objective is formulated as follows:
\begin{equation}
f^*,\{\vec{c}_{i,j}\}^* = \arg\min_{f,\{\vec{c}_{i,j}\}} \sum_{q} \sum_{d^+ \in \mathcal{D}^+_q} \sum_{d^- \in {\mathcal{D}^-_q}^\dagger} \ell(s^\dagger(q,d^+),s^\dagger(q,d^-))  
\end{equation}

\subsection{Efficiency}
The search time complexity and index size are the same as PQ. 
As we introduce PQ in Section~\ref{sec:background_pq}, the search time complexity is $O(NM+N{\rm log}n)$, and the index size is $4KD+NM \approx NM$ bytes. Compared with brute-force search, the speedup ratio is $(D+{\rm log}n)/(M+{\rm log}n)$, and the compression ratio is $4D/M$.


\begin{table*}
\small
\centering
\robustify\bfseries
\caption{
Comparison with different ANNS methods on TREC 2019 Deep Learning Track. */** denotes that JPQ performs significantly better than baselines at $p < 0.05/0.01$ level using the two-tailed pairwise t-test. }
\label{tab:results_anns}
\begin{threeparttable}	
\begin{tabular}{@{}ll
!{\color{lightgray}\vrule}
S[table-format=2.2,table-column-width=11mm]
S[table-format=1.3,table-column-width=12mm]S[table-format=1.3,table-column-width=12mm]S[table-format=1.3,table-column-width=12mm]S[table-format=1.3,table-column-width=12mm]
!{\color{lightgray}\vrule}
S[table-format=2.2,table-column-width=11mm]
S[table-format=1.3,table-column-width=12mm]S[table-format=1.3,table-column-width=12mm]S[table-format=1.3,table-column-width=16mm]@{}}
\toprule
\textbf{} & \multirow{2}{*}{\textbf{Model}} 
& {\textbf{Index}}
& \multicolumn{2}{c}{\textbf{MARCO Passage}} & \multicolumn{2}{c!{\color{lightgray}\vrule}}{\textbf{DL Passage}} 
& {\textbf{Index}}
& \multicolumn{2}{c}{\textbf{MARCO Doc}} & {\textbf{DL Doc}} \\
 &  & GB & {MRR@10} & {R@100} & {NDCG@10} & {R@100} & GB & {MRR@100} & {R@100} & {NDCG@10} \\ \midrule
\multicolumn{2}{l}{\textbf{Non-exhaust. ANNS}}   \\
 & Annoy~\cite{Github:annoy} & 30.96 & 0.150** & 0.274** & 0.431** & 0.157** & 11.36 & 0.304** & 0.638** & 0.553** \\
 & FALCONN~\cite{andoni2015practical} & {>25.34\tnote{a}} & 0.297** & 0.825** & 0.567** & 0.411** & {>9.20\tnote{a}} & 0.339** & 0.869** & 0.541** \\
 & FLANN~\cite{muja2009flann} & 32.85 & 0.319** & 0.839** & 0.607** & 0.430** & 10.60 & 0.366** & 0.895** & 0.593  \\
 & IMI~\cite{babenko2014inverted} & 25.39 & 0.331** & 0.828** & 0.610* & 0.431* & 9.24 & 0.376** & 0.869** & 0.590 \\
 & HNSW~\cite{malkov2018efficient} & 25.76 & 0.334* & 0.848** & 0.624 & 0.454 & 9.35 & 0.378** & 0.879** & 0.588* \\
\multicolumn{2}{l}{\textbf{Compressed ANNS}}   \\
 & PQ~\cite{jegou2010product} & 0.79 & 0.123** & 0.527** & 0.323** & 0.213** & 0.29 & 0.152** & 0.544** & 0.273**  \\
 & LSH~\cite{johnson2019billion} & 0.79 & 0.302** & 0.824** & 0.578** & 0.417** & 0.29 & 0.351** & 0.882** & 0.576**  \\
 & ITQ+LSH~\cite{gong2012iterative} & 0.80 & 0.296** & 0.825** & 0.582** & 0.415** & 0.29 & 0.347** & 0.874** & 0.570**  \\
 & ScaNN~\cite{guo2020accelerating} & 0.79 & 0.288** & 0.818** & 0.555** & 0.386** & 0.29 & 0.335** & 0.866** & 0.534**  \\
 & HNSW+OPQ~\cite{johnson2019billion} & 0.95 & 0.309** & 0.818** & 0.596** & 0.424** & 0.35 & 0.357** & 0.876** & 0.543** \\
 & OPQ~\cite{ge2013optimized} & 0.83 & 0.305** & 0.839** & 0.594** & 0.435** & 0.30 & 0.361** & 0.892** & 0.588**  \\
 & OPQ+ScaNN & 0.83 & 0.313** & 0.843** & 0.614** & 0.442* & 0.30 & 0.361** & 0.897** & 0.583**  \\
 & DPQ~\cite{chen2020differentiable,zhang2021joint} & 0.83 & 0.311** & 0.848** & 0.601** & 0.453 & 0.30 & 0.367** & 0.899** & 0.585**  \\
 \multicolumn{2}{l}{\textbf{Ours}}   \\
 & JPQ & 0.83 & \bfseries 0.341 & \bfseries 0.868 & \bfseries 0.677 & \bfseries 0.466 & 0.30 & \bfseries 0.401 & \bfseries 0.914 & \bfseries 0.623 \\ 
\bottomrule
\end{tabular}
\begin{tablenotes}
\footnotesize
   \item[a] FALCONN does not support saving index to disk and it is hard to infer the exact index size at run time.
\end{tablenotes}
\end{threeparttable}
\end{table*}


\section{Experimental Setup}
Here we present our experimental settings, including datasets, baselines, and implementation details.

\subsection{Datasets and Metrics}

We conduct experiments with two large-scale ad-hoc retrieval benchmarks from the TREC 2019 Deep Learning Track~\cite{craswell2020overview, bajaj2016ms}. Passage Retrieval has a corpus of $8.8$M passages, $0.5M$ training queries, $7k$ development queries~(henceforth, MARCO Passage), and $43$ test queries~(DL Passage). 
Document Retrieval has a corpus of $3.2M$ documents, $0.4M$ training queries, $5k$ development queries~(MARCO Doc), and $43$ test queries~(DL Doc). 
For both tasks, we report the official metrics and R@100 based on the full-corpus retrieval results. 

\subsection{Baselines}
We exploit six types of models as baselines, including different retrieval models and ANNS methods. 

\subsubsection{Traditional BoW Models} \mbox{}

BM25~\cite{robertson1994some} is a popular probabilistic BoW retrieval model that ranks documents based on the query terms appearing in each document. We use Anserini implementation~\cite{yang2018anserini}.

\subsubsection{Augmented BoW Models} \mbox{}

Several methods use deep language models to improve BoW models. We use them as our baselines, including doc2query~\cite{nogueira2019document}, docT5query~\cite{nogueira2019doc2query}, DeepCT~\cite{dai2019context}, and HDCT~\cite{Dai2020ContextAwareDT}.

\subsubsection{Late-Interaction Models} \mbox{}

ColBERT~\cite{Khattab2020ColBERTEA} stores the contextualized token embeddings and retrieves candidates with a late-interaction operation. 
The index size is very large because it stores token-level representations. 

\subsubsection{Brute-force DR Models} \mbox{}

Previous DR studies retrieve candidates for queries with brute-force search. We call them brute-force DR models. They share similar architectures~\cite{devlin2019bert, liu2019roberta} but differ in training process.
Baselines trained by negative sampling method include Rand Neg~\cite{huang2020embedding}, BM25 Neg~\cite{zhan2020repbert}, ANCE~\cite{xiong2020approximate}, STAR~\cite{zhan2020optimizing}, and ADORE+STAR~\cite{zhan2020optimizing}. The baseline trained by knowledge distillation is TCT-ColBERT~\cite{lin2020distilling}. 

\subsubsection{Non-exhaustive ANNS DR Models} \mbox{}

Non-exhaustive ANNS methods accelerate search but do not compress the index.
We use the following methods as baselines. Annoy~\cite{Github:annoy} is based on random projection tree forest and we set the number of trees to 100. FALCONN~\cite{andoni2015practical} is based on LSH and the recommended parameter settings are used. FLANN~\cite{muja2009flann} contains several ANNS algorithms and we use the available auto-tuning procedure to infer the best parameters. IMI~\cite{babenko2014inverted} generalizes the inverted index idea with PQ~\cite{jegou2010product} and the number of bits is set to 12. HNSW~\cite{malkov2018efficient} builds a hierarchical set of proximity graphs and we set the number of links to 8 and efconstruction to 100.

\subsubsection{Compressed ANNS DR Models} \mbox{}

Compressed ANNS DR Models compress document embeddings. 
Unsupervised compression baselines include LSH~\cite{indyk1998approximate}, ITQ+LSH~\cite{gong2012iterative}, PQ~\cite{jegou2010product}, OPQ~\cite{ge2013optimized}, ScaNN~\cite{guo2020accelerating}, OPQ+ScaNN, and HNSW+OPQ~\cite{johnson2019billion}. Note, OPQ+ScaNN uses OPQ to learn the transformation and ScaNN to compress the transformed embeddings, and HNSW+OPQ~\cite{johnson2019billion} uses HNSW to construct proximity graphs and OPQ to compress embeddings. 
The supervised compression baseline is DPQ~\cite{chen2020differentiable,zhang2021joint}. It is originally designed for word embedding compression~\cite{chen2020differentiable} and recommendation systems~\cite{zhang2021joint}. We implement it for document ranking.

\subsection{Implementation Details}
We build our models based on huggingface transformers~\cite{wolf2019huggingface} and Faiss ANNS library~\cite{johnson2019billion}.
All dual-encoders use the `bert-base'~\cite{devlin2019bert, liu2019roberta} architecture. The embedding dimension $D$ is 768, and similarity function $ \langle , \rangle$ is inner-product.
All ANNS baselines use STAR~\cite{zhan2020optimizing} as dual-encoders.
Here is how we implement JPQ.
$K$ is set to 256, and $M$ is set to 16, 24, 32, 48, 64, and 96 for experiments in different parameter settings. 
We use OPQ~\cite{ge2013optimized} to learn a linear transformation of embeddings and then use PQ~\cite{jegou2010product} for compression.
As for JPQ's training settings, we use the same parameters for both retrieval tasks.
We use AdamW optimizer, batch size of $32$, and LambdaRank~\cite{burges2010ranknet} as pair-wise loss function. Top-200 irrelevant documents are used as hard negatives. For the query encoder, learning rate is set to $5 \times 10^{-6}$. For PQ parameters, learning rate equals to $5 \times 10^{-6}$ for $M=16/24$, $2 \times 10^{-5}$ for $M=32$, and $1 \times 10^{-4}$ for $M=48/64/96$. 

\section{Experiments}
Now we empirically evaluate JPQ to address the following three research questions:
\begin{itemize}
	\item \textbf{RQ1:} Can JPQ substantially compress the index without significantly hurting the ranking performance?
	\item \textbf{RQ2:} How does JPQ perform compared with other retrieval models?
	\item \textbf{RQ3:} How do different strategies contribute to the effectiveness of JPQ?
\end{itemize}

\label{sec:experiments}


\subsection{Comparison with ANNS Methods}
\label{sec:exp_compare_anns}
\begin{figure}
    \subfloat[MS MARCO Passage]{\label{fig:psg_mrr_memory}\includegraphics[width=0.5\linewidth]{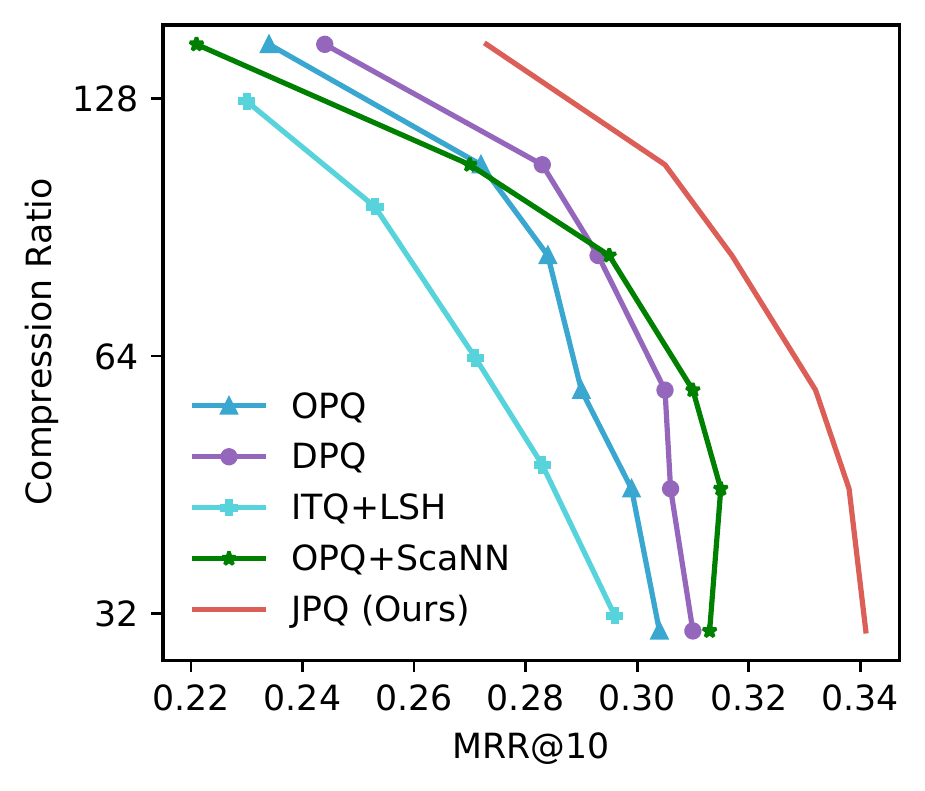}} 
    \subfloat[MS MARCO Document]{\label{fig:doc_mrr_memory}\includegraphics[width=0.5\linewidth]{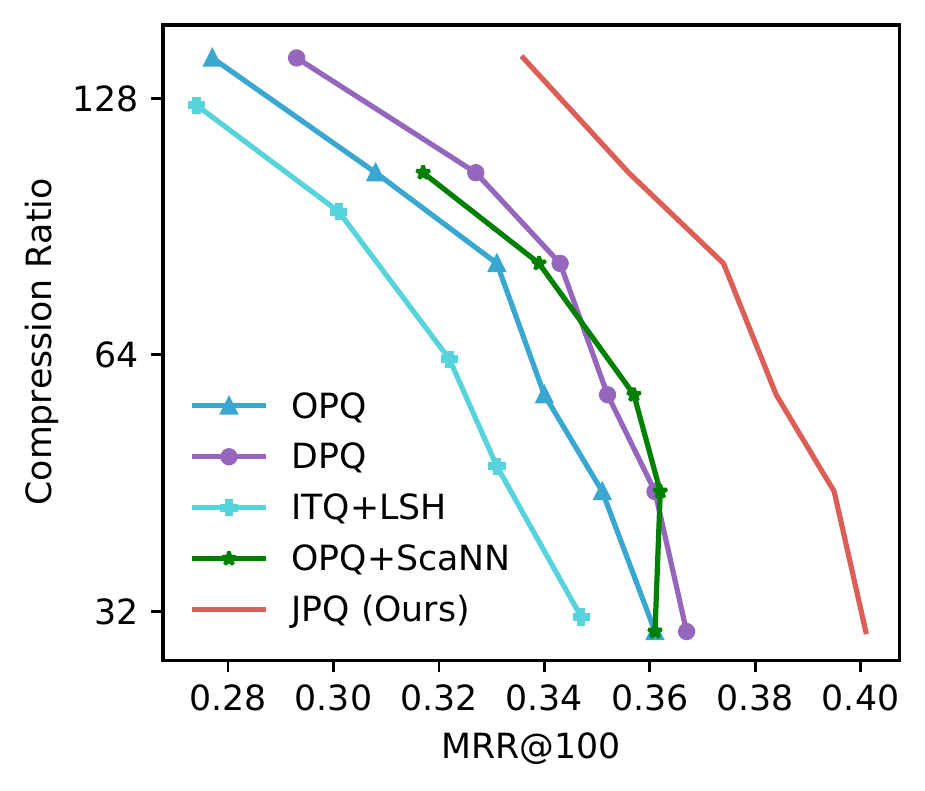}}
    \caption{Effectiveness-Memory trade-off, up and right is better.}
    \label{fig:mrr_memory}
    \vspace{-2mm}
\end{figure}

\begin{figure}
    \subfloat[MS MARCO Passage]{\label{fig:psg_mrr_speed}\includegraphics[width=0.5\linewidth]{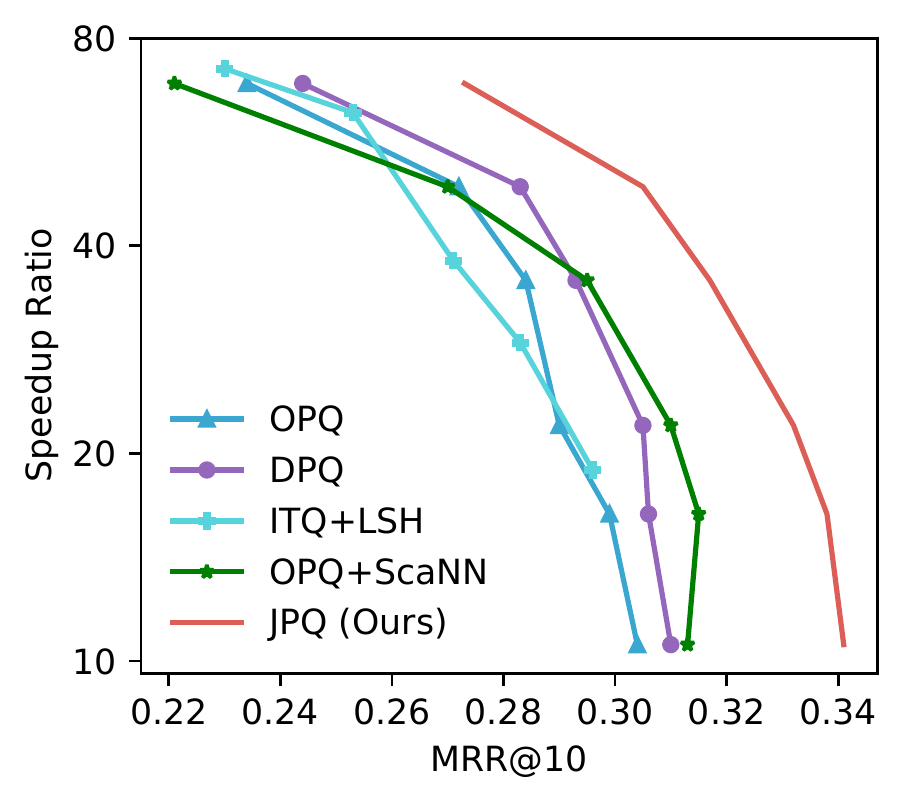}} 
    \subfloat[MS MARCO Document]{\label{fig:doc_mrr_speed}\includegraphics[width=0.5\linewidth]{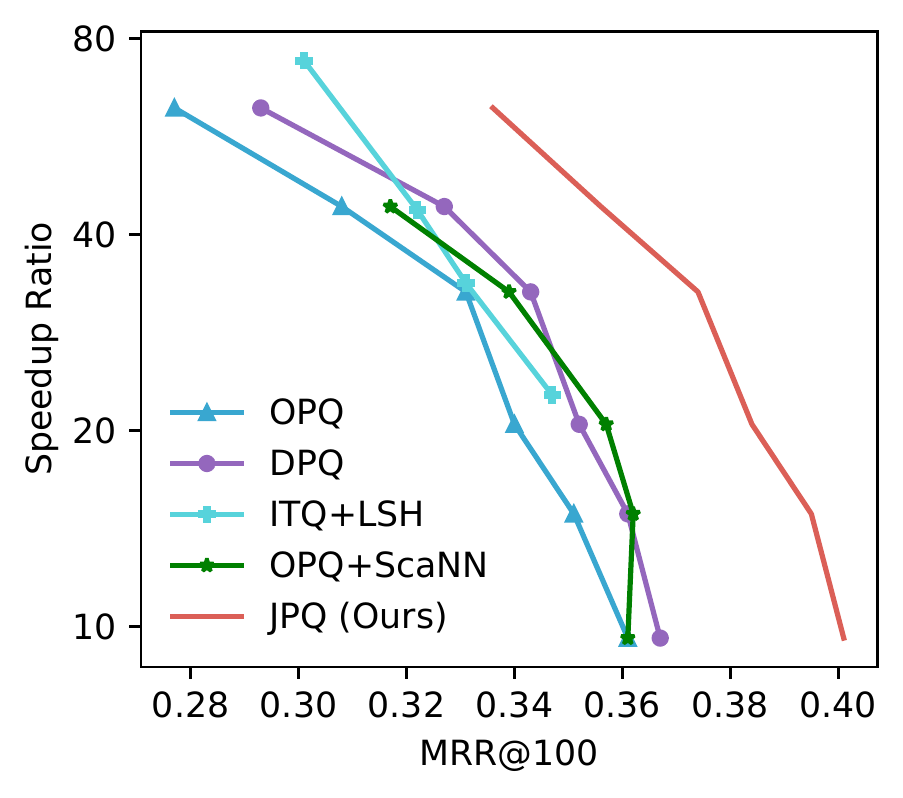}}
    \caption{Effectiveness-Speed trade-off, up and right is better.}
    \label{fig:mrr_speed}
    \vspace{-2mm}
\end{figure}

This section compares JPQ with ANNS methods to answer \textbf{RQ1}.
We utilize two types of ANNS baselines, i.e., vector compression methods and non-exhaustive ANNS methods. 
Results at a fixed trade-off setting are presented in Table~\ref{tab:results_anns}.  
Results at different trade-off settings are presented in Figures~\ref{fig:mrr_memory} and \ref{fig:mrr_speed}, which show the effectiveness-memory and effectiveness-speed curves, respectively. 
Next, we proceed to study these results in detail. 

\subsubsection{Comparison with vector compression methods} \mbox{}

Compressed ANNS baselines as well as JPQ use roughly the same compression ratio~(30x) in Table~\ref{tab:results_anns}.
According to the results, we see that JPQ leads the ranking performance by a large margin on all metrics. 
Among these baselines, DPQ~\cite{chen2020differentiable,zhang2021joint} is also a joint learning method. Although it benefits from supervised signals, it only marginally outperforms other unsupervised methods.
The problem is that it still uses reconstruction error as loss and does not design an end-to-end negative sampling method.
Conversely, JPQ adopts three strategies to improve the end-to-end ranking performance specifically, enabling it to substantially outperform both unsupervised and supervised baselines.

Now we investigate ranking performance at different compression and speedup ratios.
According to Figures~\ref{fig:mrr_memory} and \ref{fig:mrr_speed}, JPQ strongly outperforms vector compression baselines regardless of different trade-off settings. 
Moreover, results also show that JPQ is effective even with a very high compression or speedup ratio.
For example, when the compression ratio is about 140x, DPQ, the most effective baseline, incurs 28\% performance loss on the document retrieval task, whereas JPQ only incurs 15\% performance loss.

\begin{table}
\small
\centering
\robustify\bfseries
\caption{
Latency in seconds on passage and document retrieval tasks measured with one GeForce 2080Ti GPU and one Intel Xeon E5-2630 V4 CPU~(single thread). 
}
\label{tab:results_latency}
\begin{threeparttable}
\begin{tabular}{@{}ll
!{\color{lightgray}\vrule}C{0.9cm}C{0.9cm}C{0.9cm}C{0.9cm}
@{}}
\toprule
\textbf{} & \multirow{2}{*}{\textbf{Model}} 
& \multicolumn{2}{c}{\textbf{Passage}} 
& \multicolumn{2}{c}{\textbf{Document}}
 \\
 &  & CPU & GPU & CPU & GPU  \\ \midrule
\multicolumn{2}{l}{\textbf{BoW}}   \\
 & BM25~\cite{yang2018anserini}  & 0.06 & {n.a.} & 0.05 & {n.a.}  \\
 & docT5query~\cite{nogueira2019doc2query} & 0.10 & {n.a.} & 0.05 & {n.a.} \\
\multicolumn{2}{l}{\textbf{Brute-force DR}}   \\
 & ANCE FirstP~\cite{xiong2020approximate}& {\multirow{3}{*}{7.60}} & {\multirow{3}{*}{n.a.\tnote{a}}} & {\multirow{3}{*}{2.50}} & {\multirow{3}{*}{0.08}} \\
 & TCT-ColBERT~\cite{lin2020distilling} & \\
 & ADORE+STAR~\cite{zhan2021optimizing} & \\
\multicolumn{2}{l}{\textbf{Late-Interaction}}   \\
 & ColBERT~\cite{Khattab2020ColBERTEA} & {n.a.} & {0.46\tnote{b}} & {n.a.} & {n.a.} \\
\multicolumn{2}{l}{\textbf{Compressed ANNS DR}}   \\
 & OPQ~\cite{ge2013optimized} & {\multirow{2}{*}{0.73}} & {\multirow{2}{*}{0.09}} & {\multirow{2}{*}{0.28}} & {\multirow{2}{*}{0.04}} \\
 & JPQ~(Ours) & \\
\bottomrule
\end{tabular}
\begin{tablenotes}
\footnotesize
   \item[a] The passage index (26GB) is too big to fit into one 2080Ti GPU.
   \item[b] We include ColBERT’s latency reported in its paper. 
\end{tablenotes}
\end{threeparttable}
\end{table}

\begin{figure}
    \includegraphics[width=0.9\linewidth, keepaspectratio=True]{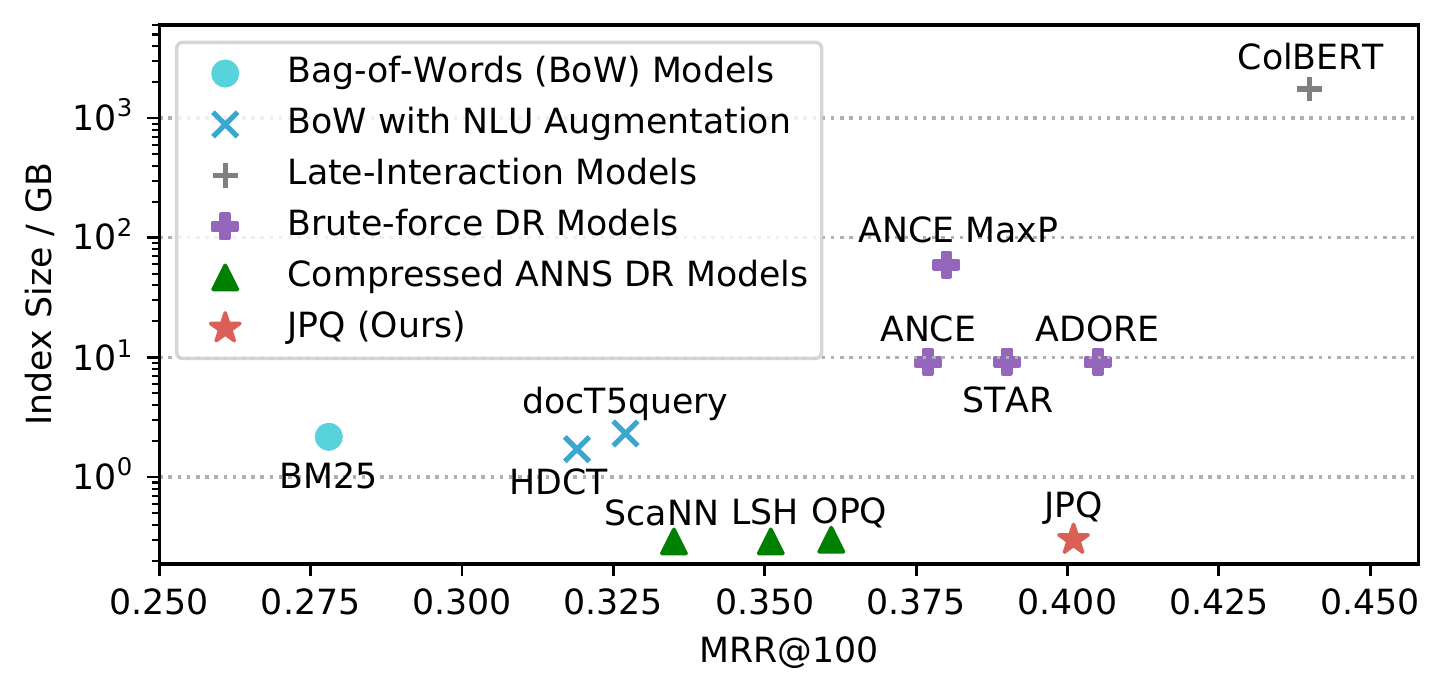}
    \caption{Effectiveness~(MRR@100) versus Index Size~(log-scale) for different retrieval methods on MS MARCO Document Ranking~\cite{bajaj2016ms}. The index size of JPQ is only 1/5833 of the size of ColBERT.
    } 
    \label{fig:ms_doc_all_models}
    \vspace{-2mm}
\end{figure}

\begin{table*}
	\small
    \centering
    \robustify\bfseries
    \caption{
    Comparison with BoW models on TREC 2019 Deep Learning Track. */** denotes that JPQ performs significantly better than baselines at $p < 0.05/0.01$ level using the two-tailed pairwise t-test. }
    \label{tab:compare_bow}
    \begin{tabular}{@{}ll
    !{\color{lightgray}\vrule}
    S[table-format=1.2,table-column-width=11mm]
    S[table-format=1.3,table-column-width=12mm]S[table-format=1.3,table-column-width=12mm]S[table-format=1.3,table-column-width=12mm]S[table-format=1.3,table-column-width=12mm]
    !{\color{lightgray}\vrule}
    S[table-format=1.2,table-column-width=11mm]
    S[table-format=1.3,table-column-width=12mm]S[table-format=1.3,table-column-width=12mm]S[table-format=1.3,table-column-width=16mm]@{}}
    \toprule
    \textbf{} & \multirow{2}{*}{\textbf{Model}} 
    & {\textbf{Index}}
    & \multicolumn{2}{c}{\textbf{MARCO Passage}} & \multicolumn{2}{c!{\color{lightgray}\vrule}}{\textbf{DL Passage}} 
    & {\textbf{Index}}
    & \multicolumn{2}{c}{\textbf{MARCO Doc}} & {\textbf{DL Doc}} \\
     &  & GB & {MRR@10} & {R@100} & {NDCG@10} & {R@100} & GB & {MRR@100} & {R@100} & {NDCG@10} \\ \midrule
    \multicolumn{2}{l}{\textbf{Traditional BoW}} \\
     & BM25~\cite{yang2018anserini} & 0.59 & 0.187** & 0.670** & 0.497** & 0.460 & 2.17 & 0.278** & 0.807** & 0.523**  \\
    \multicolumn{2}{l}{\textbf{Augmented BoW}} \\
     & doc2query~\cite{nogueira2019document} & 0.65 & 0.215** & 0.713** & 0.533** & 0.471 & {n.a.} & {n.a.} & {n.a.} & {n.a.}  \\ 
     & DeepCT~\cite{dai2019context} & 0.48 & 0.242** & 0.754** & 0.569** & 0.455 & {n.a.} & {n.a.} & {n.a.} & {n.a.}  \\ 
     & HDCT~\cite{Dai2020ContextAwareDT}  & {n.a.} & {n.a.} & {n.a.} & {n.a.} & {n.a.} &  1.71 & 0.319** & 0.843** & {n.a.} \\
     & docT5query~\cite{nogueira2019doc2query} & 0.96 & 0.272** & 0.819** & 0.642 & \bfseries 0.514 & 2.31 & 0.327** & 0.861** & 0.597  \\ 
    \multicolumn{2}{l}{\textbf{Ours}} \\
     & JPQ & 0.83 & \bfseries 0.341 & \bfseries 0.868 & \bfseries 0.677 &  0.466 & 0.30 & \bfseries 0.401 & \bfseries 0.914 & \bfseries 0.623 \\ 
    \bottomrule
    \end{tabular}
    \end{table*}

\subsubsection{Comparison with non-exhaustive ANNS methods} \mbox{}

Non-exhaustive ANNS baselines are tuned to be as fast as JPQ in Table~\ref{tab:results_anns}, specifically, 10x faster than brute-force search on CPU.
According to the results, we see that JPQ is significantly more effective in both retrieval tasks, which is a bit surprising since those baselines do not compress the index and require some memory overhead to build data structures. We believe their problem is similar to that of vector compression baselines, i.e., separation between encoding and indexing. On the contrary, JPQ benefits from end-to-end training and improves the ranking performance.

\subsection{Comparison with Retrieval Models}
This section uses existing retrieval models as baselines to answer \textbf{RQ2}.
We firstly compare JPQ with baselines in general, and then separately compare JPQ with different types of baselines in detail.

\subsubsection{Overall Comparison} \mbox{}

We summarize the ranking performance, index size, and latency of representative retrieval models in Figure~\ref{fig:ms_passage_all_models}, Figure~\ref{fig:ms_doc_all_models}, and Table~\ref{tab:results_latency}.
According to the figures, although existing neural retrieval models, i.e., brute-force DR models and late-interaction models, are more effective than BoW models, they significantly increase the index size by several orders of magnitude. When the indexes of brute-force DR models are compressed by LSH~\cite{indyk1998approximate} or OPQ~\cite{ge2013optimized}, the retrieval effectiveness is severely hurt.  
Therefore, the results seem to imply that large index sizes are necessary for high-quality ranking.

In contrast with trading index size for ranking performance, JPQ achieves high ranking effectiveness with a tiny index.
According to Figures~\ref{fig:ms_passage_all_models} and \ref{fig:ms_doc_all_models}, it outperforms BoW model by a large margin with similar or even much smaller index sizes. It gains similar ranking performance with state-of-the-art brute-force DR models while substantially compressing the index. 
As for the retrieval latency, Table~\ref{tab:results_latency} shows that JPQ is as fast as BoW models with GPU acceleration. 
Compared with brute-force DR models, JPQ provides 10x speedup on CPU and 2x speedup on GPU. 
Compared with ColBERT, JPQ provides 5x speedup. 
These results highlight the effectiveness of JPQ.

\subsubsection{Separate Comparison} \mbox{}

In the following, we use more comprehensive ranking results to separately compare JPQ with different types of retrieval models.

\begin{table}
    \centering
    \small
    \caption{Comparison between late-interaction models and JPQ+BERT two-stage methods on MARCO Passage dataset.}
    \label{tab:compare_rerank_colbert}
    \begin{tabular}{ll!{\color{lightgray}\vrule}ccc} \toprule
    & \textbf{Model} & \textbf{Latency (ms)} & \textbf{GPUs} & \textbf{MRR@10}
    \\ \midrule
    
	\multicolumn{2}{l}{\textbf{Late-Interaction}}  \\
	& {ColBERT}~\cite{Khattab2020ColBERTEA} & 458 & 1 & 0.360  \\ 

	\multicolumn{2}{l}{\textbf{JPQ+$\text{BERT}_\text{base}$}}  \\
	& {Re-rank Top 10} & 116 & 1 & 0.367 \\
	& {Re-rank Top 30} & 187 & 1 & 0.378 \\
	\multicolumn{2}{l}{\textbf{JPQ+$\text{BERT}_\text{large}$}}  \\
	& {Re-rank Top 10} & 184 & 1 & 0.372 \\
	& {Re-rank Top 30} & 406 & 1 & \textbf{0.387} \\

	\bottomrule 
    \end{tabular}
\end{table}
\begin{table*}
	\small
    \centering
    \robustify\bfseries
    \caption{
    Comparison with existing DR models and late-interaction models on TREC 2019 Deep Learning Track. }
    \label{tab:compare_dr_colbert}
    \sisetup{input-decimal-markers = .,input-ignore = {,},group-separator={,}, group-four-digits = true} %
    \begin{threeparttable}    	
    \begin{tabular}{@{}ll
    !{\color{lightgray}\vrule}
    S[table-format=1.0,table-column-width=11mm]<{x}
    S[table-format=1.0,table-column-width=12mm,retain-explicit-plus]<{\%}
    S[table-format=1.0,table-column-width=12mm,retain-explicit-plus]<{\%}
    S[table-format=1.0,table-column-width=12mm,retain-explicit-plus]<{\%}
    S[table-format=1.0,table-column-width=12mm,retain-explicit-plus]<{\%}
    !{\color{lightgray}\vrule}
    S[table-format=1.0,table-column-width=11mm]<{x}
    S[table-format=1.0,table-column-width=12mm,retain-explicit-plus]<{\%}
    S[table-format=1.0,table-column-width=12mm,retain-explicit-plus]<{\%}
    S[table-format=1.0,table-column-width=12mm,retain-explicit-plus]<{\%}L{0mm}
    @{}}
    \toprule
    \textbf{} & \multirow{2}{*}{\textbf{Model}} 
    & \multicolumn{1}{c}{\textbf{Index}}
    & \multicolumn{2}{c}{\textbf{MARCO Passage}} & \multicolumn{2}{c!{\color{lightgray}\vrule}}{\textbf{DL Passage}} 
    & \mc{\textbf{Index}}
    & \multicolumn{2}{c}{\textbf{MARCO Doc}} & \mc{\textbf{DL Doc}} \\
     &  & \mc{GB} & \mc{MRR@10} & \mc{R@100} & \mc{NDCG@10} & \mcl{R@100} & \mc{GB} & \mc{MRR@100} & \mc{R@100} & \mc{NDCG@10} \\ \midrule
    & JPQ & \mc{0.83} & \mc{0.341} &  \mc{0.868} &  \mc{0.677} &  \mcl{0.466} & \mc{0.30} &  \mc{0.401} &  \mc{0.914} &  \mc{0.623} \\ 
     \multicolumn{2}{l}{\textbf{Brute-force DR}} \\
     & Rand Neg~\cite{huang2020embedding} & +30 & -12 & -2 & -10 & -1 & +30 & -18 & -6 & -8 & \\
     & BM25 Neg~\cite{gao2020complementing} &  +30 & -9 & -6 & -10 & -22 & +30 & -21 & -13 & -13 & \\
     & ANCE FirstP~\cite{xiong2020approximate} & +30 & -1 & -1 & -4 & -5 & +30 & -6 & -2 & -2 & \\
     & ANCE MaxP~\cite{xiong2020approximate} & +30 & -1 & -1 & -4 & -5 & +196 & -5 & -1 & +2 & \\
     & TCT-ColBERT~\cite{lin2020distilling} & +30 & -2 & -1 & -1 & -2 & +196 & -17 & -5 & -2 & \\
     & STAR~\cite{zhan2021optimizing} & +30 & -0 & -0 & -5 & +0 & +30 & -3 & -0 & -3 & \\
     & ADORE+STAR~\cite{zhan2021optimizing} & +30 &  +2 & +1 & +1 & +2 & +30 & +1  & +1 & +1 & \\ 
    \multicolumn{2}{l}{\textbf{Late-Interaction}} \\
     & ColBERT~\cite{Khattab2020ColBERTEA} & \mc{186x} & \mc{+6\%\tnote{a}} & \mc{+2\%\tnote{a}} & \mc{n.a.} & \mcl{n.a.} & \mc{5833x\tnote{b}} & \mc{+10\%\tnote{b}} &  \mc{+1\%\tnote{b}} & \mc{n.a.} &\\
    \bottomrule
    \end{tabular}
    \begin{tablenotes}
	\footnotesize
   	\item[a] We include MRR@10 reported in the paper and consult the authors about R@100 due to lack of open-sourced ColBERT checkpoint, 
   	\item[b] We consult the authors about ColBERT's performance on document ranking task since it is not included in the original paper.
	\end{tablenotes}
    \end{threeparttable}
    \end{table*}

Table~\ref{tab:compare_bow} compares JPQ with traditional BoW models as well as the augmented variants. 
Results show that JPQ substantially outperforms them in terms of both accuracy~(MRR) and recall. 
For example, its index is only one-eighth the size of docT5query~\cite{nogueira2019doc2query} on the document retrieval task, and it still improves MRR@100 and R@100 by 23\% and 6\%, respectively.

Table~\ref{tab:compare_dr_colbert} shows the index size and ranking performance of brute-force DR models relative to JPQ. Even if JPQ compresses the index by 30x, it outperforms some competitive brute-force DR baselines and gains similar retrieval performance with the state-of-the-art one. Results clearly demonstrate that JPQ effectively compresses the index with only marginal ranking performance loss.

Table~\ref{tab:compare_dr_colbert} shows ColBERT's retrieval performance relative to JPQ. Besides, due to ColBERT's high latency, we add a reranking process to JPQ and show the results in Table~\ref{tab:compare_rerank_colbert}.
According to Table~\ref{tab:compare_dr_colbert}, JPQ gains similar recall even if its index size is 186x smaller on passage retrieval task and 5833x smaller on document retrieval task.
According to Table~\ref{tab:compare_rerank_colbert}, JPQ+BERT substantially outperforms ColBERT with much smaller latency.

\subsection{Ablation Study}
\label{sec:ablation_study}
\begin{figure}
    \subfloat[MRR@10]{\label{fig:ablation_mrr}\includegraphics[width=0.48\linewidth]{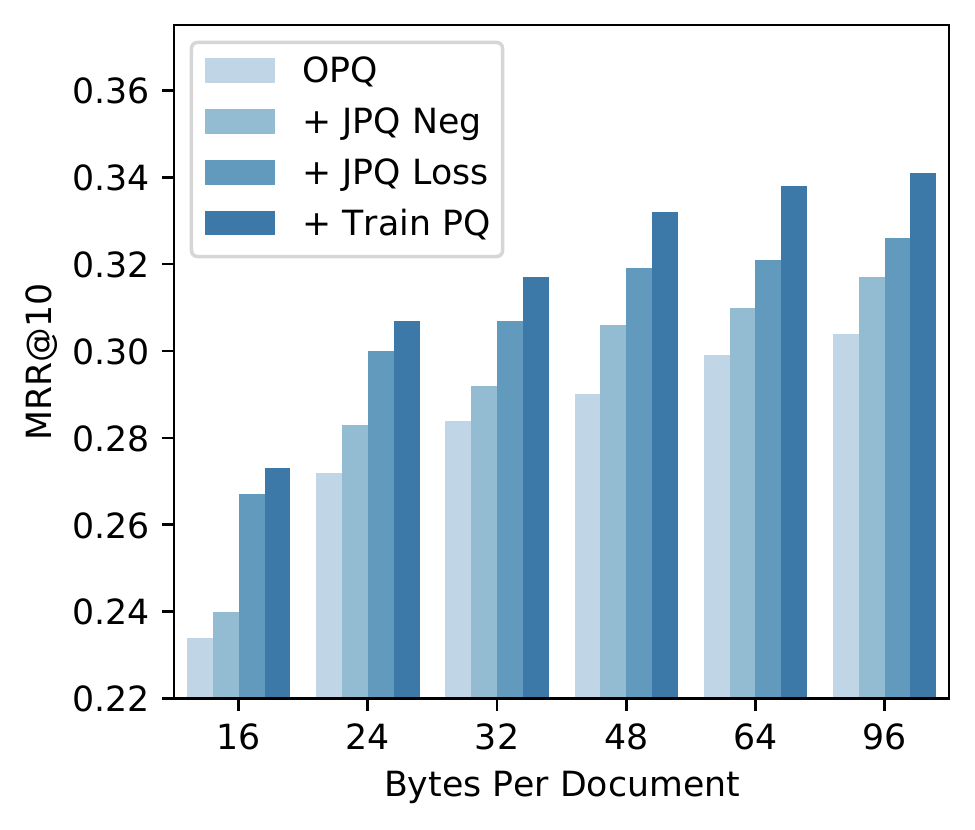}} 
    \subfloat[Recall@100]{\label{fig:ablation_recall}\includegraphics[width=0.48\linewidth]{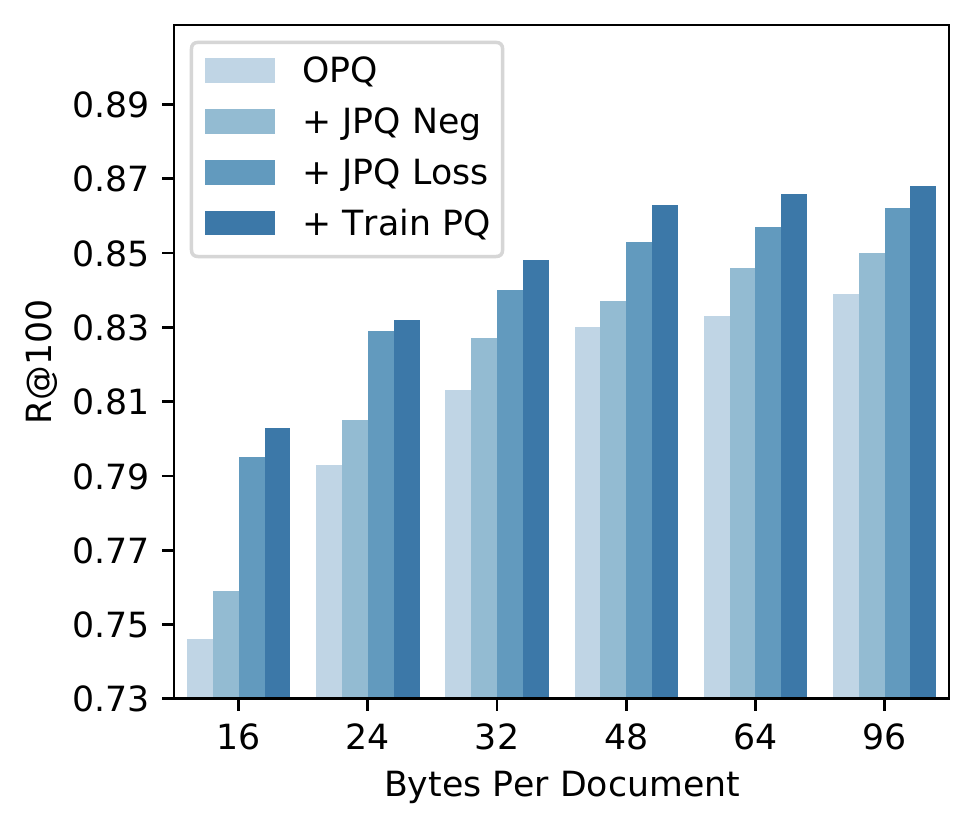}} 
    \caption{The ablation study of JPQ on MARCO Passage dataset.}
    \label{fig:ablation}
    \vspace{-2mm}
\end{figure}

JPQ employs three strategies, i.e., ranking-oriented loss, PQ centroid optimization, and end-to-end negative sampling. This section investigates their contributions to answer \textbf{RQ3}.

We conduct an ablation study on the passage retrieval task by incrementally adding the three strategies to the basic OPQ quantization method~\cite{ge2013optimized}. Specifically, we use the following four model variants:
\begin{itemize}[leftmargin=*]
	\item OPQ~\cite{ge2013optimized}: It is a popular quantization method and serves as the initialization of JPQ. 
	\item +JPQ Neg: Given OPQ initialization, it further trains the query encoder with end-to-end negative sampling. 
	\item +JPQ Loss: Based on `+JPQ Neg', it further trains the query encoder with ranking-oriented loss while the PQ index is fixed.
	\item +Train PQ: Based on `+JPQ Loss', it further trains PQ centroid embeddings. In fact, it is exactly JPQ.
\end{itemize}
We conduct experiments at different compression settings and report MRR@10 and Recall@100.
Results are shown in Figures~\ref{fig:ablation_mrr} and \ref{fig:ablation_recall}.
We can see that all three strategies contribute to the effectiveness of JPQ regardless of different parameter settings.
When fewer bytes are used to encode one document, the contribution of `ranking-oriented loss' is more prominent. We believe that fewer bytes lead to more difference between $s^\dagger(q,d)$ and $s(q,d)$, and therefore computing loss with $s^\dagger(q,d)$ is more important. 
When more bytes are used to encode one document, the contribution of `Train PQ' is more prominent, especially in terms of MRR@10. We think using more bytes to encode one document helps diversify the quantized document embeddings, and thus training centroid embeddings can better fit the dataset.

\section{Conclusions}
This paper presents JPQ, a novel framework for dense retrieval that aims to achieve high-quality ranking performance with a compact index rather than following the trend of trading index size for ranking performance. It jointly optimizes encoding and compression processes in an end-to-end manner with three carefully designed strategies, i.e., ranking-oriented loss, PQ centroid optimization, and end-to-end negative sampling. We conduct experiments on two publicly available benchmarks, where JPQ achieves impressive performance. For example, even if JPQ compresses the index by 30x and accelerates retrieval by 10x on CPU and 2x on GPU, it outperforms some competitive brute-force DR models and gains similar ranking performance with the state-of-the-art one. The results clearly demonstrate the effectiveness of JPQ and highlight that a small embedding index can still be very effective in the first-stage retrieval. 

\begin{acks}
This work is supported by the National Key Research and Development Program of China (2018YFC0831700), Natural Science Foundation of China (Grant No. 61732008, 61532011, 61902209, U2001212), Beijing Academy of Artificial Intelligence (BAAI), Tsinghua University Guoqiang Research Institute, Beijing Outstanding Young Scientist Program (NO. BJJWZYJH012019100020098) and Intelligent Social Governance Platform, Major Innovation \& Planning Interdisciplinary Platform for the "Double-First Class" Initiative, Renmin University of China.
\end{acks}

\clearpage
\bibliographystyle{ACM-Reference-Format}
\balance
\bibliography{references}

\end{document}